\documentclass[aps,prd,superscriptaddress,final,twocolumn,groupedaddress]{revtex4}

\usepackage{graphicx,nico}
\usepackage{multirow} 
\usepackage{color}
\usepackage[normalem]{ulem}

\begin{document}

\preprint{Edinburgh 2011/25, SHEP-1125}

\title{Opening the Rome-Southampton window for operator mixing matrices}

\newcommand\edinb{SUPA, School of Physics, The University of
  Edinburgh, Edinburgh EH9 3JZ, UK}
\newcommand\soton{School of Physics and Astronomy, University of
  Southampton,  Southampton SO17 1BJ, UK}
\newcommand\columb{Physics Department, Columbia University, New York, NY 10027, USA. }

\author{R.~Arthur}\affiliation{\edinb}
\author{P.A.~Boyle}\affiliation{\edinb}
\author{N.~Garron}\affiliation{\edinb}
\author{C.~Kelly}\affiliation{\columb}

\author{A.T.~Lytle}\affiliation{\soton}
\collaboration{RBC and UKQCD Collaborations}

\date{Dated: December 25, 2011}

\begin{abstract}
We show that the running of operators which mix under renormalization can be computed
fully non-perturbatively as a product of continuum step scaling matrices.
These step scaling matrices are obtained by taking the ``ratio'' of Z matrices computed 
at different energies in an RI-MOM type scheme for which twisted boundary conditions
are an essential ingredient.
Our method allows us to relax 
the bounds of the Rome-Southampton window.
We also explain why such a method is important in view of the light quark 
physics program of the RBC-UKQCD collaborations.
To illustrate our method, using $n_f=2+1$ domain-wall fermions, 
we compute the non-perturbative running matrix 
of four-quark operators needed in $K\to\pi\pi$ decay and neutral kaon mixing. 
Our results are then compared to perturbation theory.

\end{abstract}


\maketitle

\noindent {\bf Introduction}\\
Lattice QCD has now reached a stage of high precision in 
flavour physics and plays a crucial r\^ole 
in the quest for new physics.
Precision phenomenology requires using a non-perturbative scheme 
to renormalize quantities obtained from lattice simulations,
avoiding ill convergent and low order lattice perturbation theory.
One possibility is to use the Schr\"odinger functional (SF) scheme, 
a theoretically appealing method which allows for a smooth 
connection between low energy - where the hadronic matrix elements are computed 
on the lattice - and very high energy, where the perturbative series is accurate
\cite{Luscher:1992an,Sint:1993un}.
This connection is done through the use of step-scaling 
functions~\cite{Luscher:1991wu,Luscher:1992zx,Luscher:1993gh}.
Unfortunately, the range of operators actually computed in the SF scheme is rather limited;
perhaps this is due to the fact that a peculiar perturbation theory
(different from the infinite volume one) is required. 
Another possibility - very popular in the lattice community - is to use a RI-MOM type 
scheme~\cite{Martinelli:1994ty}:
such a scheme is  theoretically sound, 
relatively easy to implement and the connection to $\msbar$ or any other perturbative scheme
is done using continuum infinite volume perturbation theory. 
Thus one benefits from the recent multi-loop computations 
achieved by several groups (see for example~\cite{Gorbahn:2010bf,Almeida:2010ns,Gracey:2011fb}).
In an RI-MOM type scheme 
one numerically computes the off-shell amputated vertex function 
$G_{\rm O}$ of the operator of interest $O$ between external states 
with given momenta and fixed gauge gluonic configurations. 
We project this quantity onto its Dirac-colour structure and 
obtain the quantity $\Lambda_{\rm O}$ (the choice of projector is
in general not unique and the precise definition of the scheme
depends on this choice of projector). 
We take all the quark masses to be degenerate and equal to $m$.
Then one requires that in the chiral limit 
the renormalized projected-amputated operator matches its tree-level value~\cite{Martinelli:1994ty},
i.e. if $a$ is the lattice spacing, $n$ is the number of quark fields, $\psi$, 
in the operator and $\mu$ is the renormalization scale
(which depends on the choice of external momenta), 
and $F$ is the tree level value of $\Lambda_{\rm O}$,
one imposes  \\
\begin{equation}
\label{eq:Zdef}
{Z_{O}^{\schemeS}(\mu,a)}\;{Z^{-n/2}_{\rm \psi}(\mu,a)} 
\lim_{m\to 0}{\Lambda_{\rm O}(\mu,a,m)} = F \;.
\end{equation}
In this way one obtains, non-perturbatively, the scheme and scale dependent 
renormalization factors $Z_{O}^{\schemeS}(\mu,a)$. 
The precise definition of the lattice scheme $\schemeS$ depends 
on the details of the implementation. 
In the second step, one converts the result to a scheme more appropriate for  
phenomenological applications.
In the case of an effective low-energy theory one can express the Hamiltonian
by a sum of local operators multiplied by some Wilson coefficients. 
One must find a scheme 
where both the renormalized matrix
elements of these operators and the corresponding Wilson coefficients
can be computed. 
Typically, one matches the Z factors obtained in the lattice scheme at $\mu\sim 2-3$ GeV
to $\overline{\rm MS}$. 
Equivalently, (if perturbation theory is accurate) one can divide the vertex function computed in 
the lattice scheme by the corresponding running to obtain the renormalization group 
invariant Z factor and then convert to the desired scheme and scale.
It is important to note that the running and 
the matching are usually computed only perturbatively
(we discuss some recent implementations of a non-perturbative running in the next section).

In order to keep the discretisation effects under control, and at the same time
access a region where perturbation theory can be applied,
RI-MOM schemes require the existence of the so-called Rome-Southampton window:
ideally one would require  $\Lambda_{\rm QCD} \ll \mu \ll a^{-1}$. 
In practice, this window can be quite narrow or even closed
and most of the lattice computations are done with $a\mu \sim 1$.
In fact the size of the discretisation effects depends on the lattice action, 
and it is usually admitted that for an off-shell O(a)-improved action 
it is sufficient to require $(a\mu/\pi)^2 \ll 1$.
Concerning the other side of the window the situation 
was greatly improved when it was realised that 
a different choice of kinematics, called non-exceptional, 
strongly suppresses some infrared contributions to the vertex 
function~\cite{Aoki:2007xm,Sturm:2009kb}.
Nevertheless, this window is still an important limitation,
in particular when considering light pion masses 
where the physical volume - and thus the lattice spacing - has to be large.
This could become a real issue when one tries to tackle the very challenging computation of
$K\to\pi\pi$ decays because one needs even 
larger physical volumes.
A new step scaling method has recently been introduced~\cite{Arthur:2010ht} 
that allows to address the window problem for RI-MOM. 
This method has already been applied for multiplicatively renormalized 
operators~\cite{Aoki:2010pe,Durr:2010aw}.
\vspace{0.5cm}

\noindent {\bf Strategy} \\
In this work we apply the method presented in~\cite{Arthur:2010ht}, 
generalised to the case 
where several operators mix under renormalization.
The main idea is to separate the scale and the lattice spacing(s)
at which the bare matrix elements are computed from the ones at which the running and 
the conversion to the perturbative scheme are performed. 
Let us consider the case of four-quark operators mixing:
following again~\cite{Martinelli:1994ty}, we define a matrix of amputated, projected vertex functions
$\Lambda_{ij}=  P_j\{G_{O_i}\}$, 
where the $O_i$ form a basis under renormalization and 
the $P_i$ projects and traces onto the Dirac-colour structure of the 
corresponding operator $O_i$. 
If we denote by $F$ the tree-level value of $\Lambda$,
equation~(\ref{eq:Zdef}) becomes
\be
\label{eq:Zmat}
{Z^{\schemeS}_{ij}(\mu,a)\over {Z_{\rm A}}(a)^2} \times
\lim_{m\to 0} { \Lambda_{jk}(\mu,a,m) \over { {\Lambda_{\rm A}(\mu,a,m)}^2}}
=F_{ik}\;,
\ee
where, for convenience, we use the vertex function of the axial current
$\Lambda_{\rm A}$ to fix $Z_{\psi}$.
Equation~(\ref{eq:Zmat}) defines $Z^{\schemeS}_{ij}$ where the scheme ${\schemeS}$
depends on the choice of kinematics and projectors of the four-quark vertex function
and on the choice of quark wave function renormalization~\cite{Aoki:2010pe}.
At finite lattice spacing $a$ and for a given renormalization scale $\mu$ we consider the matrix 
\be
R_{\schemeS}(\mu,a)=\lim_{m\to 0}\left[\Lambda^2_{\rm A}(\mu,a,m) \, \Lambda^{-1}(\mu,a,m)\right]
\;,
\ee
and we define the step scaling matrix by
\be
\label{eq:ssm}
\sigma^\schemeS(\mu,s\mu)=\lim_{a\to 0}\Sigma^\schemeS(\mu,s\mu,a)=\lim_{a\to 0}{\left[{R_\schemeS}(\mu,a)R_\schemeS^{-1}(s\mu,a)\right]}\;.
\ee
We also note that $Z_{\rm A}(a)$ cancels out in the ratio since, to a very good approximation,
it does not depend on the scale $\mu$. 
One important point is that although the quantities $\Lambda$, $Z$ and $R$
depend on the details of the computation this is not the case for the 
step scaling matrix which has well-defined continuum limit 
and depends only on the choice of renormalization scheme $\schemeS$
(and on the number of flavours).
In this work we use a scheme which is called $(\gamma_\mu,\gamma_\mu)$-scheme 
in~\cite{Aoki:2010pe}: it involves non-exceptional kinematics with 
a symmetric point. 
Such a choice greatly suppresses unwanted infrared effects.\\
The strategy that we are proposing can be summarised in the following way:
\begin{itemize}
\item Consider a set of simulations with a rather large volume  
of spatial extent $L_0$ 
where physical pion masses can be simulated
and with moderately large lattice spacings $a_0$, 
but small enough for the Symanzik expansion to converge.
Compute the bare matrix elements of interest 
$\langle {\cal O}^{\rm bare}(a_{\rm 0}) \rangle $ and renormalize them
in a lattice scheme $\schemeS$ at the low energy scale $\mu_0$,
i.e. compute 
$\displaystyle\langle {\cal O}^{\schemeS}(\mu_{\rm 0}) \rangle=
\lim_{a_0 \to 0} {\left[ Z_{\cal O}^{\schemeS}(\mu_{\rm 0},a_0) \langle {\cal O}^{\rm bare}(a_0) \rangle \right]}$.
Of course the scale $\mu_{\rm 0}$ should be 
such that the associated discretisation errors are small
but, compared to the Rome-Southampton window, we do not require the non-perturbative effects to be small.
Instead one just has to ensure that the finite volume effects are negligible, 
so the renormalization window becomes 
$$L_0^{-2} \ll \mu_0^2 \ll (\pi/a_0)^2\;.$$  
\item Iterate the following step, with $\rm i={1,2,\ldots,n}$:
consider a set of simulations, with a physical volume of space extent $L_{\rm i}<L_{\rm i-1}$ 
and a set of lattice spacings $a_i$. 
With the requirement that, on each lattice, 
$$L_{\rm i}^{-2} \ll \mu^2_{\rm i-1} < \mu^2_{\rm i} \ll (\pi/a_{\rm i})^2 $$
is satisfied, 
compute the step scaling matrix by evaluating eq.~(\ref{eq:ssm}).
\item At a scale $\mu_{\rm n}$ high enough to apply perturbation theory,  
multiply by the perturbative matching and running factors corresponding 
to the desired scale and scheme (typically $\overline{\rm MS}$ at $\mu=2$ or $\mu=3$ GeV).
In this volume the usual Rome-Southampton condition holds
$$
\Lambda_{\rm QCD}^{2} \ll \mu^2_{\rm n} \ll (\pi/a_{\rm n})^2 \;.
$$
\end{itemize}

\noindent In summary, the general equation 
can be written as 
\begin{widetext}
\begin{equation}
\label{eq:main}
\langle O^{\overline{\rm MS}}(\mu) \rangle
=
C^{\overline{\rm MS}\leftarrow \schemeS}(\mu) \times
U^{\schemeS}(\mu,\mu_{\rm n}) \times
\underbrace{\sigma^{\schemeS}(\mu_{\rm n},\mu_{\rm n-1})}_{\mbox{Fine lattices} }
\times \sigma^{\schemeS}(\mu_{\rm n-1},\mu_{\rm n-2})
\times \ldots \times \sigma^{\schemeS}(\mu_{\rm 1},\mu_0)
\times \underbrace{\langle O^{\schemeS}(\mu_0) \rangle}_
{\mbox{Coarse lattices }}
\;,
\end{equation}
\end{widetext}
where $C^{\overline{\rm MS}\leftarrow \schemeS}(\mu)$ represents the matrix of matching factors which converts
the $Z$ matrix computed in the scheme $\schemeS$ to the scheme $\overline{\rm MS}$, and 
$U^{\schemeS}(\mu,\mu_{\rm n})$ is the usual running matrix in the scheme $\schemeS$ computed in
perturbation theory. 
In the previous equation we have made explicit the fact that our method consists 
in re-expressing the running matrix, usually computed in perturbation theory,
by a product of continuum non-perturbative step scaling matrices.\\

One notices that in the first $\rm n-1$ steps, the lower bound of the Rome-Southampton window ($\Lambda_{\rm QCD}$) is replaced 
by a more advantageous one ($L_i^{-1}$). In other words: we do not need to be in the perturbative regime.
Moreover a better control of the upper bound is achieved by taking the continuum limit of the step scaling function
(in particular there is no discretisation error of order $a_0^2\mu_{\rm n}^2$  
in contrast to the ``naive'' RI-MOM implementation). 
A couple of remarks are in order:
\begin{itemize}
\item 
Using twisted boundary conditions circumvents the Fourier mode constraints and 
allows us to fix the orientation of the momentum
while changing its magnitude. As emphasised in~\cite{Arthur:2010ht}, thanks to this property 
we can compute the vertex functions for an arbitrary number of points lying on the same 
scaling trajectory.
The continuum limit of the vertex function as a function of $\mu^2$ is then properly defined 
and in particular we do not need to perturbatively subtract any lattice artefact. 
The continuum extrapolation is also easier 
since, 
if we know the lattice spacings with sufficient precision, 
we can simulate any arbitrary momentum. 
This is also useful because 
our method requires that each momentum $p=\mu_{\rm 0},\mu_1,\ldots\mu_{\rm n}$ has to be common 
to different sets of lattices.
\item Since we impose periodic boundary conditions, one could worry about 
the use of perturbation theory 
in a small volume, where the space extent is of the order of (or smaller than) 
$\Lambda_{\rm QCD}^{-1}$~\cite{GonzalezArroyo:1981vw}. 
However we claim that in a non-exceptional graph 
with hard external momenta, decoupling will ensure, for 
$\mu\gg L^{-1}$, that our computation is free from finite volume effects
and finite volume perturbation  theory is not needed.
While this might need further investigation, it is not  
relevant here since we consider only ``infinite'' volumes (of spatial extent much larger 
than $\Lambda_{\rm QCD}^{-1}$). 
\end{itemize}

Before closing this section we wish to mention
other works on non-perturbative running in RI-MOM.
Taking the continuum limit of the ratio of $Z$ factors at different energies 
in an RI-MOM scheme was first proposed 
in~\cite{Donini:1999sf}
but the authors did not address how to match the momenta computed with
different lattice spacings such that the lattice artefacts have an $a^2$ 
expansion.
Zhestkov~\cite{Zhestkov:2001hx} used fine tuning of $\beta$ in the quenched approximation
to exactly match the Fourier modes but did not define a quantity 
which has a well-defined continuum limit.
More recent work has looked at the ratios of $Z$s~\cite{Constantinou:2010gr,Durr:2010aw,Durr:2011ap}.
However, how to continuum extrapolate the distinct Fourier modes with distinct lattice artefacts
must be addressed.  
Some cases have model input like perturbative subtraction of the lattice discretisation 
effects, or rules of thumb such as use of $\sin(p)$ instead of $p$. 
Instead here we follow~\cite{Arthur:2010ht} and implement twisted boundary conditions 
to keep the orientation of the momenta fixed and give the observables a smooth $a^2$ dependence,
and allow for a full non-perturbative continuum extrapolation.
\vspace{0.5cm}

\noindent {\bf Renormalization of kaon weak matrix elements.} \\
In this section we give the definitions of the kaon four-quark
operators that we consider in this work. 
We refer the reader who would like to find more details about this part
to the recent reviews~\cite{Sachrajda:2011tg, Lellouch:2011qw}.\\

\begin{figure}[!t]
\begin{center}
\includegraphics[width=5cm]{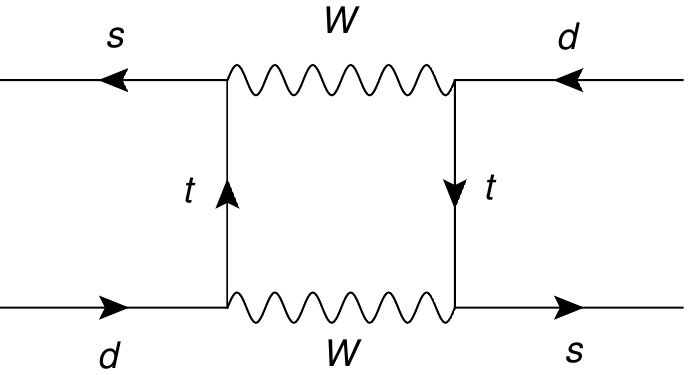}
\caption[]{Example of box diagram contributing to $K^{0}-\overline{K^{0}}$ mixing in the Standard model.} 
\label{fig:box}
\end{center}
\end{figure}

In the standard model, neutral kaon mixing is dominated 
by box diagrams like the one shown in figure~\ref{fig:box}.
The non-perturbative contributions are given by 
$\langle \overline K^0 | O^{\Delta s=2}_{\rm VV+AA} | K^0 \rangle $, 
where $O^{\Delta s=2}_{\rm VV+AA}$ is the parity conserving part 
of $(\overline s \gamma_\mu(1-\gamma_5) d)(\overline s \gamma_\mu(1-\gamma_5) d)$.
It is well known that this operator belongs to the (27,1) representation
of $SU(3)_{\rm L} \times SU(3)_{\rm R}$ and renormalizes multiplicatively.
To study neutral kaon mixing beyond the standard model it is useful to introduce
the so-called SUSY basis of $\Delta s=2$ operators.
In this basis $O_1^{\Delta s=2}$ is the standard model operator, $O_i^{\Delta s=2}, \, i>1$
are beyond the standard model (BSM) operators. 
Denoting by $\alpha$ and $\beta$ the colour indices,
one has
\bea
(27,1) &
\left.
\;\;
\begin{array}{ccc}
O_1^{\Delta s=2}&=&
(\overline s_\alpha \gamma_\mu (1-\gamma_5) d_\alpha)\,
(\overline s_\beta  \gamma_\mu (1-\gamma_5) d_\beta)\,,\\
\end{array}
\right.
\;\;
\nn\\
(6,\overline 6) &
\left\{
\begin{array}{lll}
O_2^{\Delta s=2}&=&
(\overline s_\alpha (1-\gamma_5) d_\alpha)\,
(\overline s_\beta  (1-\gamma_5) d_\beta)\,,\\
\label{eqO3}
O_3^{\Delta s=2}&=&
(\overline s_\alpha  (1-\gamma_5) d_\beta)\,
(\overline s_\beta   (1-\gamma_5) d_\alpha)\,,\\
\end{array}
\right.
\nn\\
(8,8) & 
\left\{
\begin{array}{rcl}
O_4^{\Delta s=2}&=&
(\overline s_\alpha  (1-\gamma_5) d_\alpha)\,
(\overline s_\beta   (1+\gamma_5) d_\beta)\,,\\
\label{eqO5}
O_5^{\Delta s=2}&=&
(\overline s_\alpha  (1-\gamma_5) d_\beta)\,
(\overline s_\beta   (1+\gamma_5) d_\alpha)\,.
\end{array}
\right.\nn
\eea
The operators have been studied with various lattice 
formulations, see for example~\cite{Babich:2006bh, Wennekers:2008sg, Dimopoulos:2010wq}.
As we wrote explicitly in the previous equations, 
$O_2^{\Delta s=2}$ and $O_3^{\Delta s=2}$ transform like 
$(6,\overline 6)$ under $SU(3)_{\rm L} \times SU(3)_{\rm R}$
and then mix together under renormalization.
Likewise $O_4^{\Delta s=2}$ and $O_5^{\Delta s=2}$ belong to $(8,8)$
and also mix together. 
Thus in a scheme which preserves chiral symmetry the five-by-five renormalization
matrix is block diagonal:  
the only non-zero $Z$ factors can be divided in three subgroups: 
a single factor for the $(27,1)$
operator and two, two-by-two matrices for the BSM operators.
In practice it is convenient 
to work in another basis where all the operators are colour unmixed
and we consider only the parity even component of the four-quark operators.
Using the notation 
$\Gamma \otimes \Gamma \rightarrow (\overline s_\alpha \Gamma d_\alpha)\,
(\overline s_\beta  \Gamma  d_\beta)$ 
we define the renormalization basis by:
\bea
(27,1) & &
\left.
\;\;
\begin{array}{ccc}
Q_1^{\Delta s=2} &=& \gamma_\mu \otimes \gamma_\mu +\gamma_\mu \gamma_5 \otimes  \gamma_\mu \gamma_5 \,,
\end{array}
\right.
\;\;
\nn\\
(8,8) & &
\left\{
\begin{array}{lll}
Q_2^{\Delta s=2}  &=& {\gamma_\mu \otimes \gamma_\mu -\gamma_\mu \gamma_5 \otimes  \gamma_\mu \gamma_5} \,, \\
Q_3^{\Delta s=2}  &=& {{\rm I} \otimes {\rm I} - \gamma_5 \otimes \gamma_5 } \,,
\end{array}
\right.
\nn\\
(6,\overline 6) & &
\left\{
\begin{array}{rcl}
Q_4^{\Delta s=2} &=& {{\rm I} \otimes {\rm I} + \gamma_5 \otimes \gamma_5 } \,, \\
Q_5^{\Delta s=2} &=& \sigma_{\mu\nu} \otimes \sigma_{\mu\nu}  \,.
\end{array}
\right.\nn
\eea
The explicit relations between the two bases 
are given in the appendix.
We denote by $Z^{\Delta s=2}$ the (block diagonal) renormalization matrix 
defined in the renormalization basis.\\
It is interesting to note that 
the renormalization factors of some $\Delta s=1$ operators which appear in
$K\to\pi\pi$ decays can be obtained from those of the $\Delta s=2$ operators
mentioned above. 
At low energy in the $\Delta I=3/2$ channel they are three operators 
that contribute: a $(27,1)$ called ${Q'}^{\Delta s=1, \Delta I=3/2}_1$ 
which renormalizes multiplicatively and 
two $(8,8)$ which mix together: the electroweak penguins ${Q'}^{\Delta s=1, \Delta I=3/2}_{7,8}$.
We give the explicit form of these operators in the appendix.
Denoting by ${\cal Z}^{\Delta s=1}_{ij}, \quad i,j=1,7,8$ the corresponding renormalization
factors, the relation to the $Z^{\Delta s=2}_{ij}$ reads
$$
\begin{array}{c c c c c c c}
& & \quad {\cal Z}^{\Delta s=1}_{11} &=& Z^{\Delta s=2}_{11}  \,,& & \\
{\cal Z}^{\Delta s=1}_{77} &=& Z^{\Delta s=2}_{22}      \,, & \quad & {\cal Z}^{\Delta s=1}_{78} &=& -{1 \over 2} \, Z^{\Delta s=2}_{23} \,,\\
{\cal Z}^{\Delta s=1}_{87} &=& -2 \, Z^{\Delta s=2}_{32}\,, & \quad & {\cal Z}^{\Delta s=1}_{88} &=& Z^{\Delta s=2}_{33} \,.
\end{array}
$$
In this work we give some results for 
the non-perturbative running of $Q_{i=1,\ldots5}^{\Delta s=2}$ and 
${Q'}_{1,7,8}^{\Delta s=1, \Delta I=3/2}$. 
A full computation of $K\to\pi\pi$ decays requires also the renormalization 
of $\Delta I=1/2$ four-quark operators which transform like $(8,1)$ under
$SU(3)_{\rm L} \times SU(3)_{\rm R}$. Because for these operators one has to compute 
disconnected diagrams we do not consider them in this work and leave them 
for the future.
Although these operators are important in a full computation of $K\to\pi\pi$ amplitudes
the main purpose of this paper is to give a method for a computation 
of a non-perturbative running in the operators mixing case.
Furthermore the operators that we consider here already have an important phenomenological 
relevance since they allow for the computation of standard model and beyond the standard model neutral kaon mixing
matrix elements and of $K\to\pi\pi$ amplitude in the $\Delta I=3/2$ channel.
\vspace{0.5cm}\\

\noindent {\bf Numerical application}\\
The RBC-UKQCD collaboration has recently performed a computation of 
$K\to\pi\pi$ decay amplitudes (in both isopsin channels) 
with a two-pion final state~\cite{Blum:2011pu}. 
The physical matrix elements are computed from the Euclidean 
ones by using the Lellouch-L\"uscher formula~\cite{Lellouch:2000pv}.
The results are very promising but, since it was the first computation of its kind, 
an unphysical pion mass of $m_\pi\sim 422 \,\rm{MeV}$ was used. 
In order to simulate the physical kinematics the collaboration is currently repeating
the computation on a much larger volume - of spatial extent $L_0\sim 4.6\,\fm$ -
with a nearly physical pion mass of $m_\pi\sim \,140 \;\MeV$. 
Some promising preliminary results of the  $K\to\pi\pi$ matrix elements in the $\Delta I=3/2$ 
channel have been reported in~\cite{Goode:2011kb}.
The gauge action used is a modification of the Iwasaki gauge action
following the lines of~\cite{Renfrew:2009wu,Vranas:2006zk}, 
that we call the ``dislocation suppressing determinant ratio''.
\begin{figure}[!t]
\begin{center}
\includegraphics[width=8.5cm]{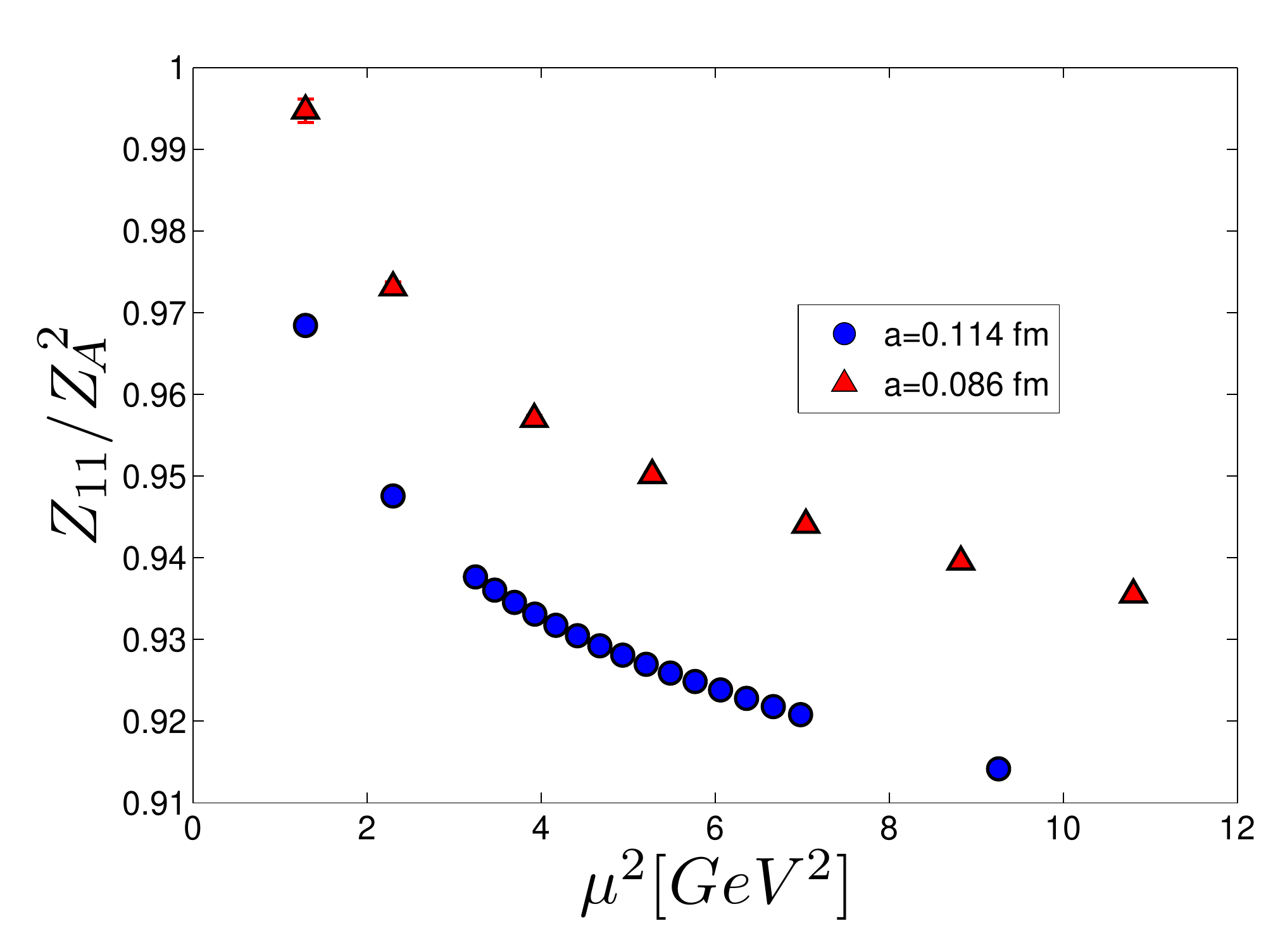}
\caption[]{Renormalization factor 
$Z_{11}/Z_{\rm A}^2=Z_{\rm B_{\rm K}}$, obtained at two different 
values of the  lattice spacing. 
Except for a few points, the error bars are smaller than the symbols.
}
\label{fig:z11}
\end{center}
\end{figure}
On the same lattice, a computation of 
the matrix element 
$\langle \overline K^0 | O^{\Delta s=2}_{\rm VV+AA} | K^0 \rangle $
is also on the way.
With a lattice spacing $a_0\sim0.14\,\fm$ 
one might worry about the size of the discretisation effects 
for a momentum of two or three GeV 
(indeed one might doubt the existence of the Rome-Southampton window
on this lattice). 
The way out is to follow the strategy explained in the previous section 
with $n=1$ (two different physical volumes).
Firstly, in the volume $L_0$ we compute the bare matrix elements 
and the renormalization factors at a low energy $\mu_0$,
where we use $\mu_0\sim 1.5 \,\GeV$.
Then the continuum limit of the step scaling matrix $\sigma(\mu_1,\mu_0)$,
with $\mu_1\in[\mu_0,\, 3 \, \rm GeV]$
is obtained from another set of simulations. 
There we use $L_1\sim \, 2.7\,\fm$
and two finer lattice spacing 
$a_1\sim \, 0.086 \,\fm$ and $0.114\,\fm$
(more details about these simulations can be found in 
\cite{Aoki:2010dy,Allton:2008pn}).
Finally we plan to combine the two results 
in order to compute the renormalized matrix element at 
$2$ or $3\GeV$ and then apply the perturbative matching 
to $\overline {\rm MS}$.
As mentioned above the two different volumes 
use two different gauge actions 
but the results can be combined together because we
extrapolate the step scaling matrix to the continuum.
Although in principle we would like to have a third lattice spacing in order to 
have a better control on the continuum extrapolations in the volume $L_1$, 
in practice the discretisation effects on the step scaling matrix elements 
appear to be small. 
Obviously this method can be applied to all sort of different 
quantities that one would like to extract from our large lattice $L_0$.
In the near future, when the next generation of supercomputers  
will be available, we plan to add a finer lattice on the 
large volume $L_0$.\\

In the rest of this section we present our results for the step
scaling matrices $\sigma(\mu_0,\mu)$ with $\mu$ varying in a range
$[\mu_0,\sim3\,\GeV]$. 
The computation of the renormalization factors on the fine lattice
is done 
using the same setup as in~\cite{Boyle:2011kn,Aoki:2010pe},
to which we refer the reader for a more detailed explanation.
The computation of the Z factors has been 
already presented in~\cite{Boyle:2011kn}.
One of the main features of our computation is the use of the
Domain-Wall fermion \cite{Kaplan:1992bt,Shamir:1993zy,Furman:1994ky}
which exhibits an almost exact chiral-flavour
symmetry. As a consequence the renormalization 
pattern is the same as in the continuum (up to some numerically 
irrelevant lattice artefacts).
Some other interesting aspects of our calculation are 
the use of the volume sources~\cite{Gockeler:1998ye} 
giving us a very good statistical precision, 
non-exceptional kinematics~\cite{Aoki:2007xm,Sturm:2009kb} 
to suppress the Goldstone pole contributions:
here we use a scheme which is called $(\gamma_{\mu},\gamma_{\mu})$-scheme 
in~\cite{Aoki:2010pe}.
In figure~\ref{fig:z11} we show the $Z$ factor
of the $(27,1)$ operator, normalized by $Z_{\rm A}^2$ and extrapolated to the chiral limit, 
obtained on the two different lattices
at the simulated momenta
(covering a range from $\sim 1.1 \,\GeV$ to $\sim 3.5\,\GeV$).
The corresponding step scaling function is computed according to the definition 
eq~(\ref{eq:ssm}).
In figure~\ref{fig:cl_s11} (left) we plot the result at finite lattice spacing, together with the 
continuum extrapolation. 
In order to determine the vertex function at any given momentum
for each ensemble we fit our data as a function of $p^2$ and
interpolate. 
As explained earlier in the text, 
because we are using twisted boundary conditions that
keep the momentum orientation fixed with respect to the lattice axes
this is a smooth interpolation. 
Having obtained in this way 
the vertex function on a number of lattice spacings at a fixed physical $p^2$ 
we extrapolate this to the continuum limit using a constant plus $O(a^2)$ ansatz for each
value of the momentum. This ansatz is justified in our approach because
as we pick a fixed momentum orientation with respect to the lattice
axes, $O(4)$ breaking lattice artefacts are well parametrised~\cite{Arthur:2010ht}. 
As one can see on the plot, 
the lattice spacing dependence is very well under control. 
In~\cite{Boyle:2011kn} we have shown that the renormalization pattern is 
the same as in the continuum, the chirally forbidden renormalization factors
being zero within error.
The chiral extrapolation is done using three different quark masses
and we found a very mild quark mass dependence for all our quantities. 
In our setup the sea quark mass of the strange is fixed (to its physical value),
whereas the valence sea quarks masses are equal to the sea light quark masses
and then extrapolated to zero. As a consequence our results are affected by a small 
systematic error, which was evaluated in~\cite{Aoki:2010pe} for the $(27,1)$ operator.
We have checked that taking the chiral extrapolation of the ratio of $Z$
gives the same result as taking the ratio of the chiral extrapolation of 
$Z$. 
In figure~\ref{fig:cl_s11} (right) we compare our non-perturbative result with the next-to-leading order (NLO)
running~\cite{Aoki:2010pe}.
We note that the running is quite small ($\sim4\%$ between $\mu\sim 1.5\,\GeV$
and $\sim 3.5\,\GeV$), while NLO perturbation theory predicts $\sim 7\%$.
With our very small error bars such a the difference is clearly visible.
In figure~\ref{fig:sigmaDs2}, we plot our results for the two (continuum) two-by-two 
step scaling matrices $\sigma(\mu,\mu_0)$ computed in the renormalization basis. 
By definition, at the matching point $\mu=\mu_0$ the matrices are equal 
to the identity.
In figure~\ref{fig:sigmaDs1} 
we compare our results for 
the electroweak penguins in the $\Delta s=1$ basis
to the next-to-leading order (NLO) running recently computed in~\cite{Lehner:2011fz}.
In Figure~\ref{fig:sigma11_over_pt} and~\ref{fig:sigmaDs1_over_pt}
we set $\mu_0=3\GeV$ and plot the non-perturbative running divided by the NLO 
prediction.
In general we find that when NLO perturbation theory is available 
the results agree qualitatively with the non-perturbative ones.
It is interesting to note that for both the $(8,8)$ and the $(6,\bar 6)$,
we found that one off-diagonal matrix element has a very small non perturbative 
running, and that the corresponding one loop anomalous dimension 
is either zero (for $\sigma_{32}$) or a very small number (for $\sigma_{45}$).

\vspace{0.5cm}
\noindent{\bf Conclusion}\\
We have presented a general method to compute
the non-perturbative continuum running 
in the operator mixing case. In particular 
this method allows the use of an RI-MOM type scheme
on a rather coarse lattice.  
We have computed this running between $\sim 1.5\, \GeV$ and $\sim 3.5 \,\GeV$ 
in the case of four-quark operators which occur in 
neutral kaon mixing (including the BSM ones) and $\Delta I=3/2$ $K\to\pi\pi$ decays.
Although the strategy we have presented is very general we have shown 
why it is important for the light quark physics program of the RBC-UKQCD collaboration
and in particular for a full computation of $K\to\pi\pi$ amplitudes.
\vspace{0.5cm}\\

\noindent{\bf Acknowledgements}\\
We thank our colleagues in the RBC and UKQCD collaborations, 
in particular Norman~Christ, Christoph~Lehner, 
and Chris~Sachrajda for suggestions and stimulating discussions. 
The calculations reported here were performed on the QCDOC computers~\cite{Boyle:2003mj,Boyle:2005gf} at
Columbia University, Edinburgh University, and at Brookhaven National Laboratory (BNL), 
Argonne Leadership Class Facility (ALCF) BlueGene/P resources at Argonne
National Laboratory (ANL), and also the resources of the STFC-funded DiRAC facility.
We wish to acknowledge support from STFC grant ST/H008845/1, DOE grant DE-FG02-92ER40699
and to EU grant 238353 (STRONGnet). R.A. is supported by SUPA prize studentship.

%
\newpage
\begin{widetext}
\begin{center}
\begin{figure}[!h]
\begin{tabular}{cc}
\includegraphics[width=9cm]{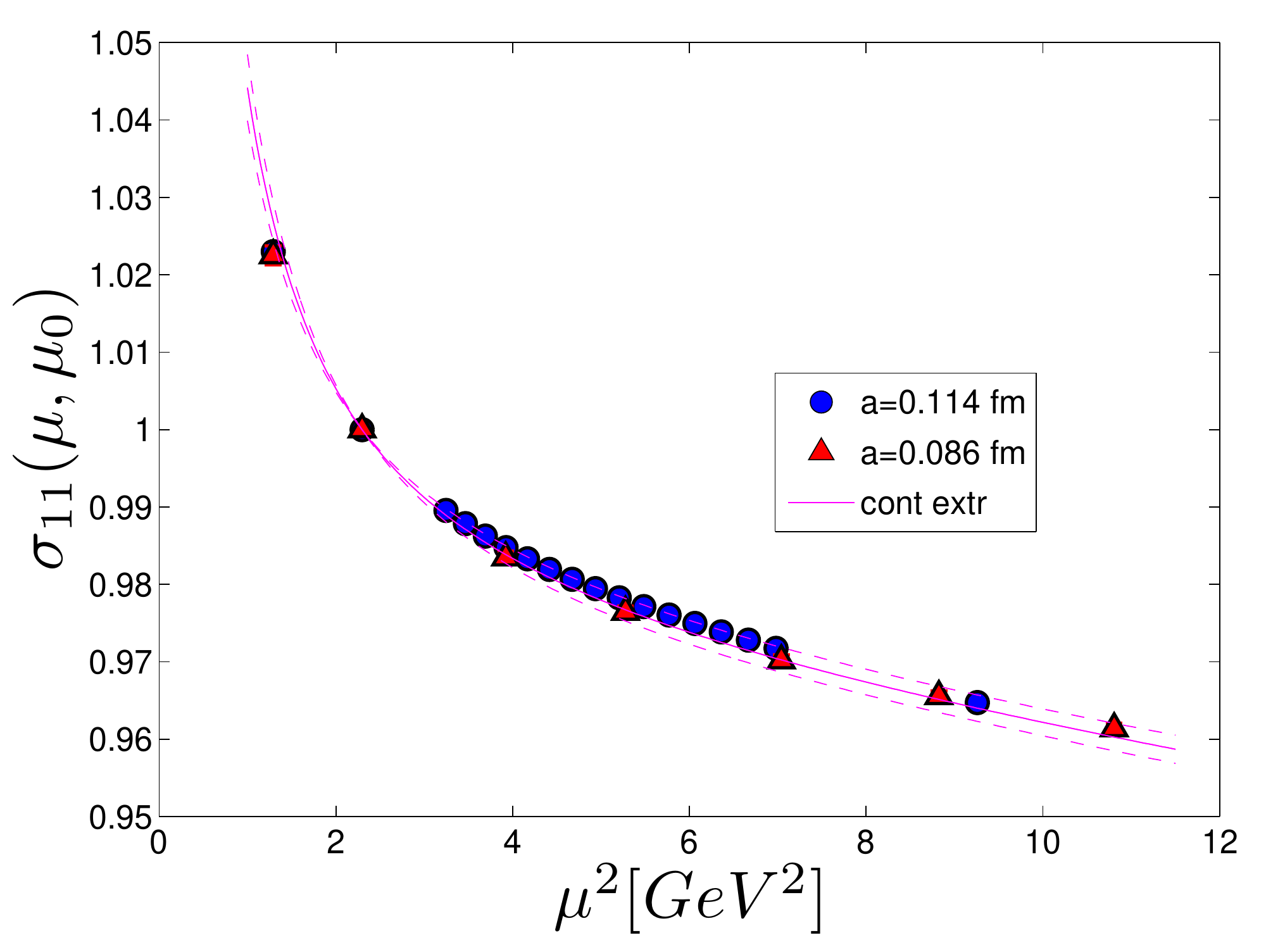}&
\includegraphics[width=9cm]{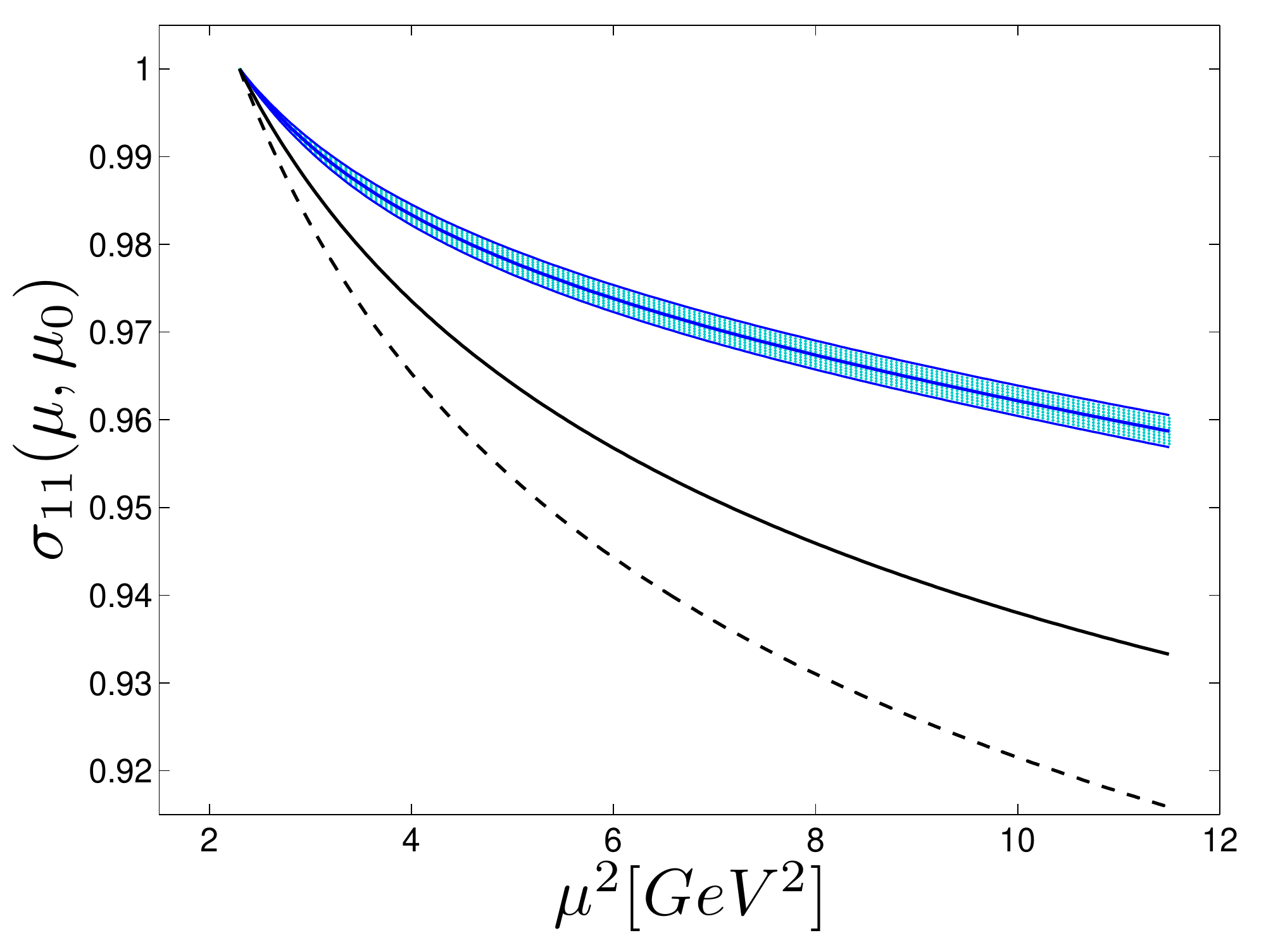}
\end{tabular}
\caption[]{Left: 
step scaling function (non-perturbative running) of the $(27,1)$ operator at finite lattice spacings and in the continuum
in the lattice SMOM $(\gamma_\mu,\gamma_\mu)$-scheme .
The energy scale $\mu_0$ is fixed to $\mu_0\sim 1.5\,\GeV$, and $\mu$ varies in the range $\left[1. \,\GeV,\, \sim 3.5\,\GeV \right]$.
The errors bars are smaller than the symbols.
Right: the same quantity is compared to perturbation theory (dashed black curve : one loop, solid black curve: two loops, 
solid blue curve: continuum extrapolation of non-perturbative running with its error) in the range $\left[\mu_0,\sim3.5\,\GeV \right]$.
}
\label{fig:cl_s11}
\end{figure}
%
%
\begin{figure}[!h]
\begin{tabular}{cc}
\includegraphics[width=9cm]{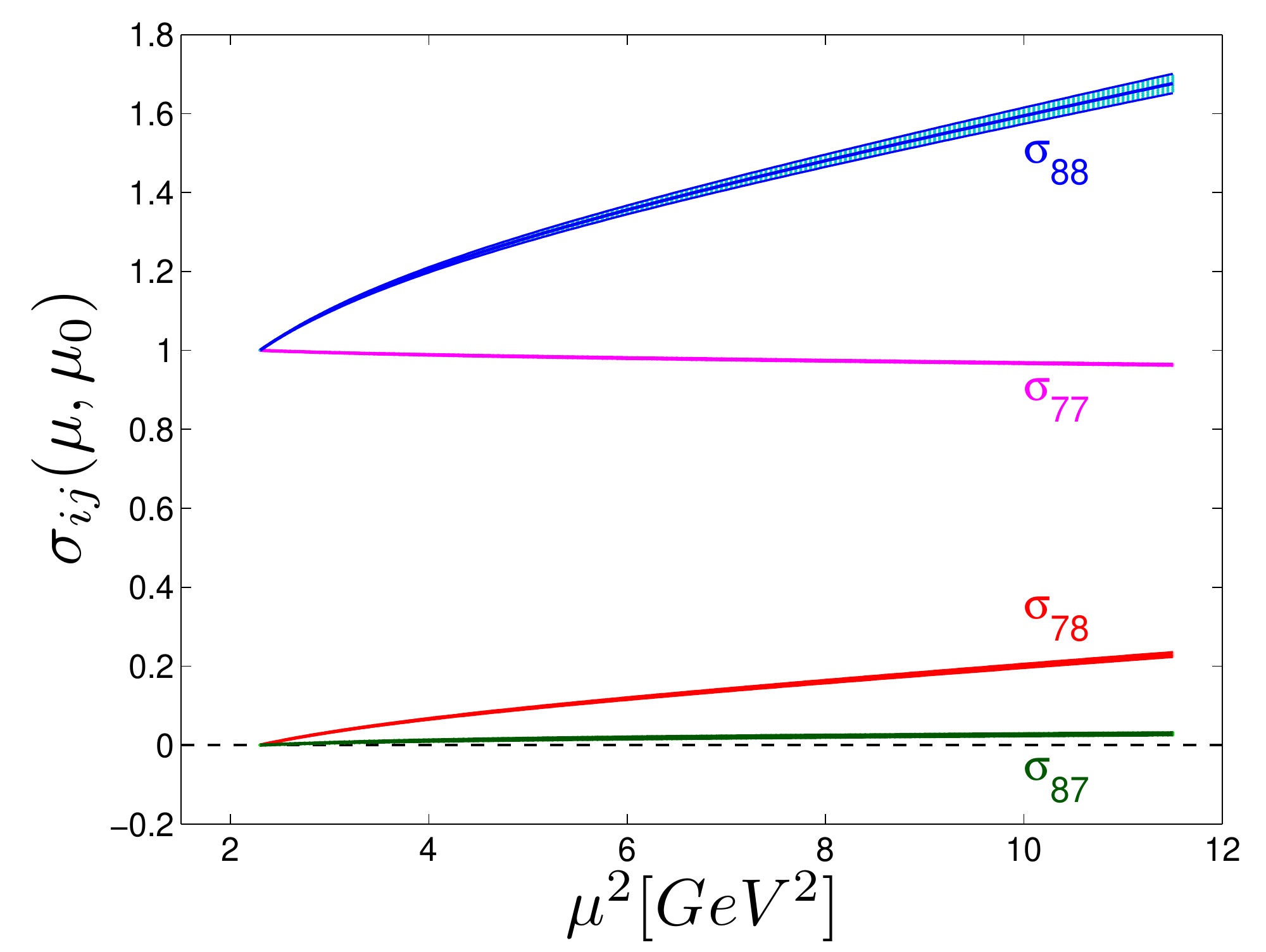}&
\includegraphics[width=9cm]{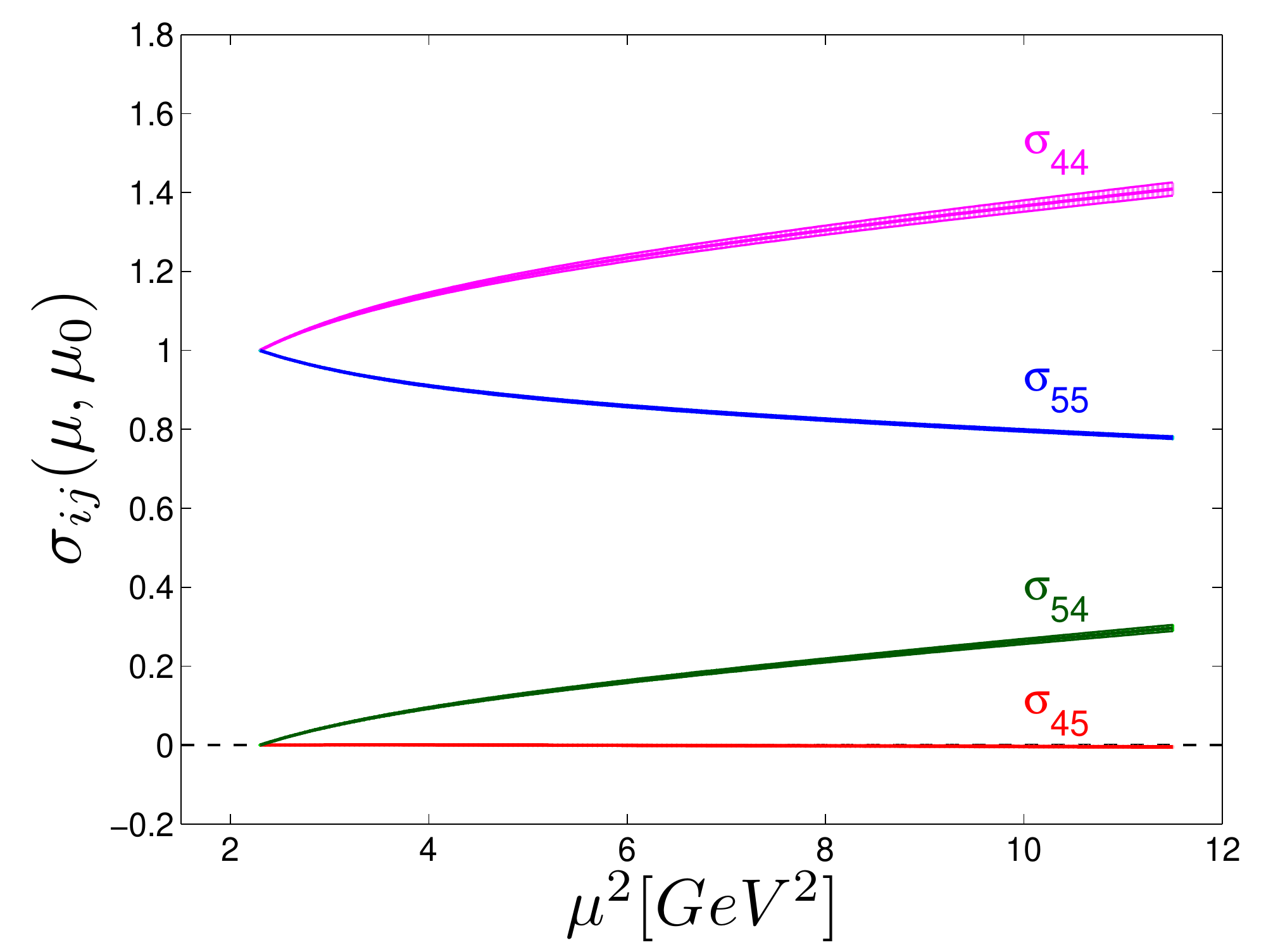}
\end{tabular}
\caption[]{
Step scaling matrix  $\sigma(\mu,\mu_0)$ given in the $\Delta s=2$ renormalization basis.
On the left panel we show the matrix elements of the $(8,8)$ operators
and on the right 
the $(6,\bar6)$ operators.
Like in figure~\ref{fig:cl_s11}, the energy scale $\mu_0$ is fixed to $\mu_0\sim 1.5\,\GeV$, 
and $\mu$ varies in the range $\left[\mu_0,\, \sim 3.5\,\GeV \right]$.
}
\label{fig:sigmaDs2}
\end{figure}
\end{center}
\end{widetext}

\appendix*
\section{}
{\bf 1.} Denoting by the superscript ``+'' the parity even component, 
the explicit relation between the two $\Delta s=2$ basis is 
\bea
(27,1) & &
\left.
\;\;
\begin{array}{ccc}
{\left[O_1^{\Delta s=2}\right]}^+ &=& Q_1^{\Delta s=2} \,,
\end{array}
\right.
\;\;
\nn\\
(6,\overline 6) & &
\left\{
\begin{array}{lll}
{\left[O_2^{\Delta s=2}\right]}^+ &=& Q_4^{\Delta s=2} \,, \nonumber\\
{\left[O_3^{\Delta s=2}\right]}^+ &=& -{1\over 2} ( Q_4^{\Delta s=2} - Q_5^{\Delta s=2}) \,,\nonumber 
\end{array}
\right.
\nn\\
(8,8) & &
\left\{
\begin{array}{rcl}
{\left[O_4^{\Delta s=2}\right]}^+ &=& Q_3^{\Delta s=2} \,, \nonumber \\
{\left[O_5^{\Delta s=2}\right]}^+ &=& -{1\over 2} Q_2^{\Delta s=2} \nonumber \,.
\end{array}
\right.\nn
\eea
%
\clearpage
\newpage
\begin{widetext}
{\bf 2.} To define the $\Delta I=3/2$ part of the $\Delta s=1$ operators, 
we follow the conventions of \cite{Blum:2001xb}
\bea
Q'_1 &=&
 (\bar s_\alpha \gamma_\mu (1-\gamma_5) d_\alpha) 
\big[ 
(\bar u_\beta  \gamma_\mu(1-\gamma_5) u_\beta) - (\bar d_\beta  \gamma_\mu(1-\gamma_5) d_\beta)) 
\big]
+( \bar s_\alpha \gamma_\mu (1-\gamma_5) u_\alpha)(\bar u_\beta \gamma_\mu (1-\gamma_5) d_\beta)) \\
Q'_7 &=& 
(\bar s_\alpha \gamma_\mu (1-\gamma_5) d_\alpha) 
\big[ 
(\bar u_\beta  \gamma_\mu(1+\gamma_5) u_\beta) - (\bar s_\beta  \gamma_\mu(1+\gamma_5) s_\beta)) 
\big]
+( \bar s_\alpha \gamma_\mu (1-\gamma_5) u_\alpha)(\bar u_\beta \gamma_\mu (1+\gamma_5) d_\beta))\\
Q'_8 &=& 
(\bar s_\alpha \gamma_\mu (1-\gamma_5) d_\beta) 
\big[ 
(\bar u_\beta  \gamma_\mu(1+\gamma_5) u_\alpha) - (\bar s_\beta  \gamma_\mu(1+\gamma_5) s_\alpha)) 
\big]
+( \bar s_\alpha \gamma_\mu (1-\gamma_5) u_\beta)(\bar u_\beta \gamma_\mu (1+\gamma_5) d_\alpha))
\eea
%
\begin{figure}[!h]
\begin{tabular}{cc}
\includegraphics[width=9cm]{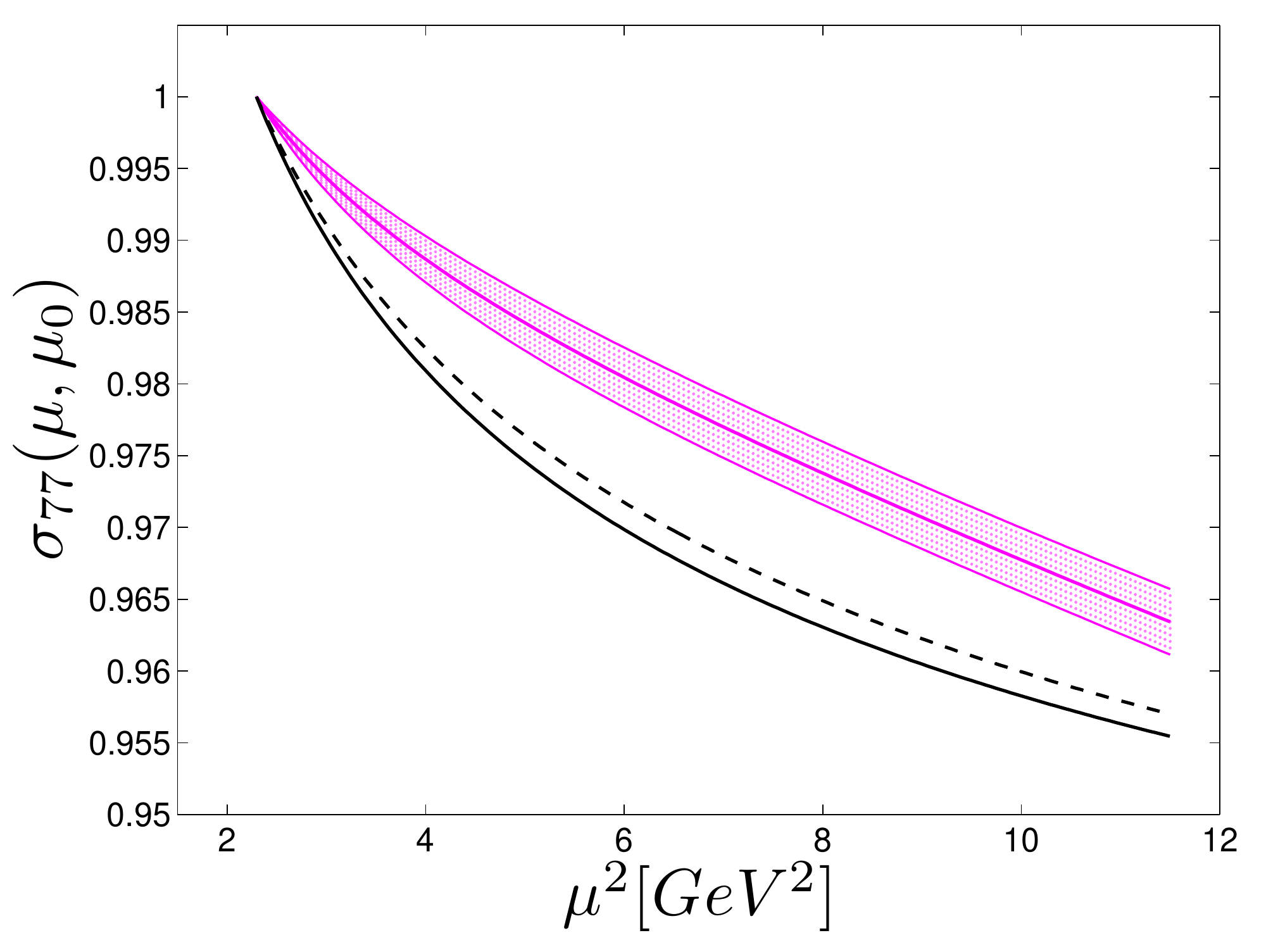}&
\includegraphics[width=9cm]{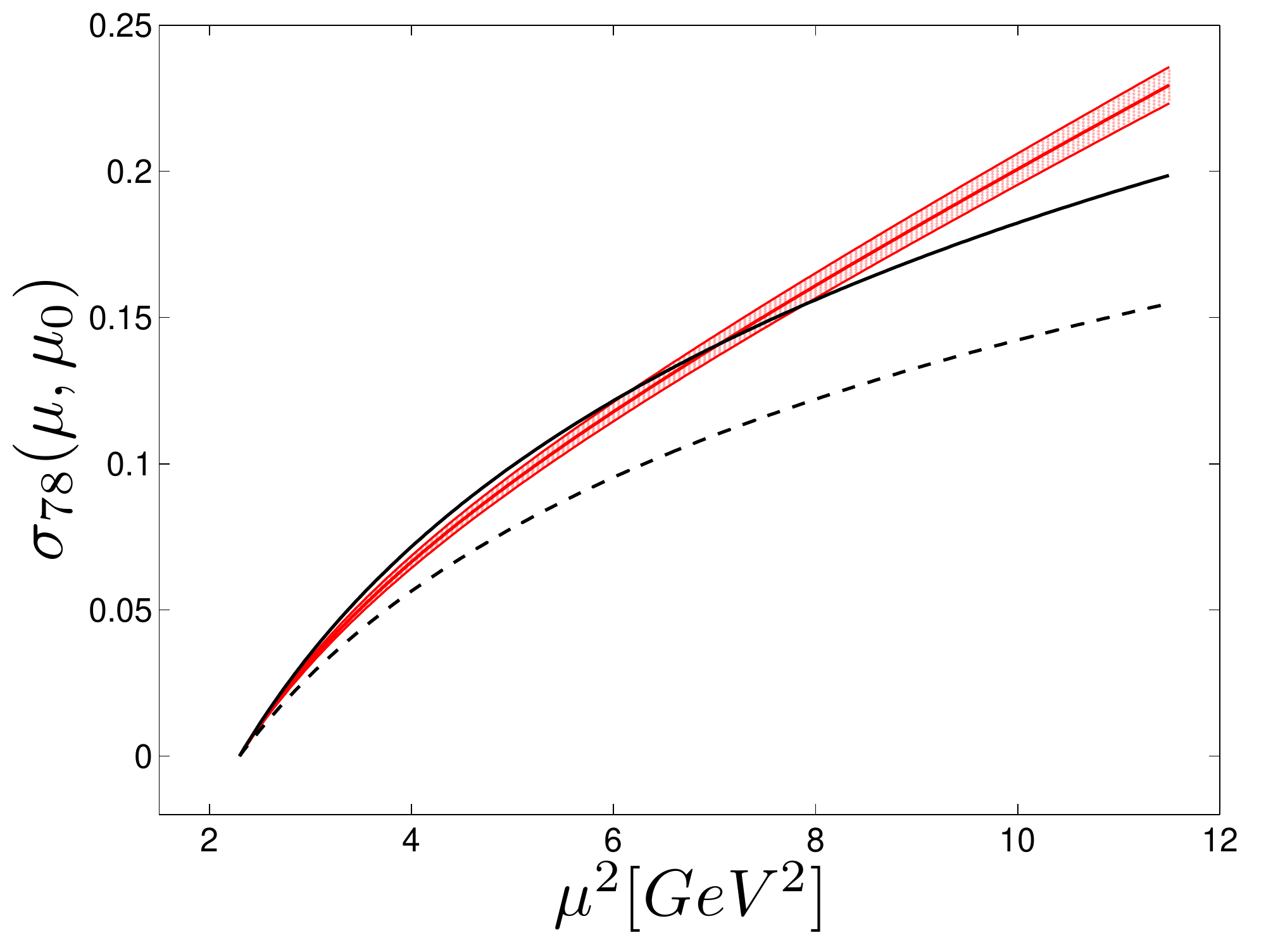}\\
\includegraphics[width=9cm]{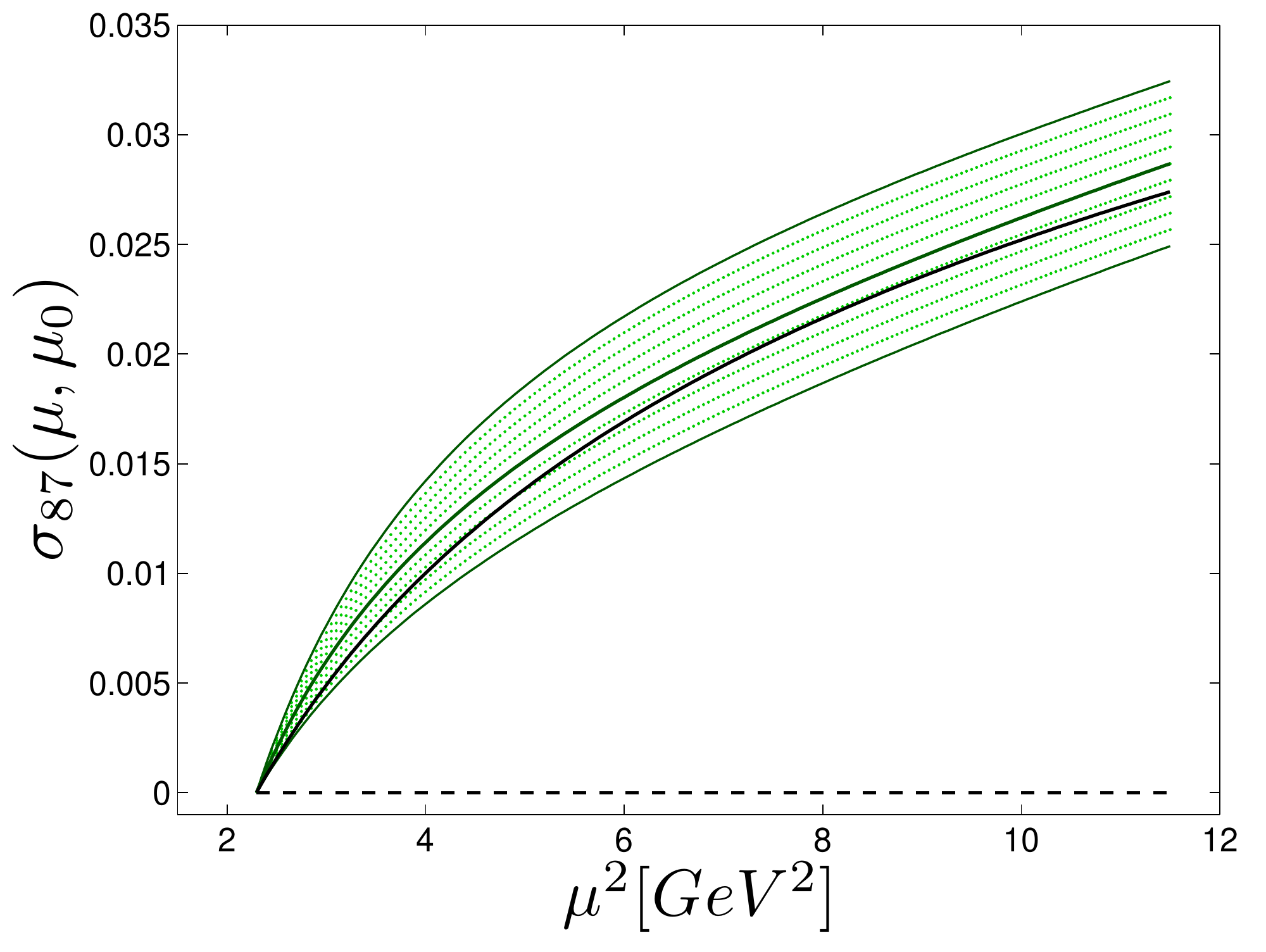}&
\includegraphics[width=9cm]{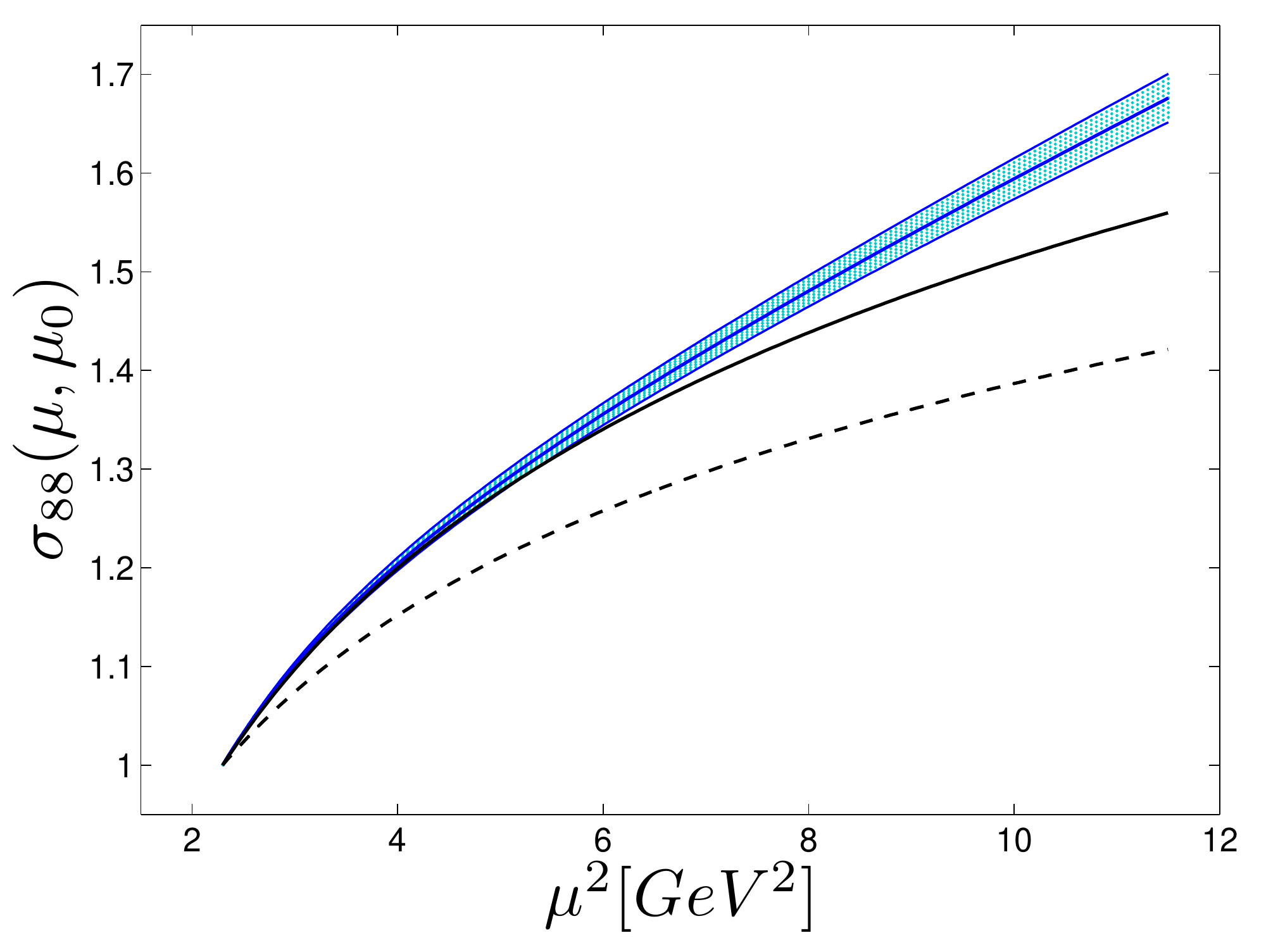}
\end{tabular}
\caption[]{
Step scaling matrix  $\sigma(\mu,\mu_0)$ of the $(8,8)$ electroweak penguins 
in the $\Delta s=1$ renormalization basis.
For each matrix element we compare to perturbation theory
(dashed black curve : one loop, solid black curve: two loops,
solid coloured curve: continuum extrapolation of non-perturbative
running with its error).
}
\label{fig:sigmaDs1}
\end{figure}
\clearpage
\newpage
\begin{figure}[!h]
\includegraphics[width=9cm]{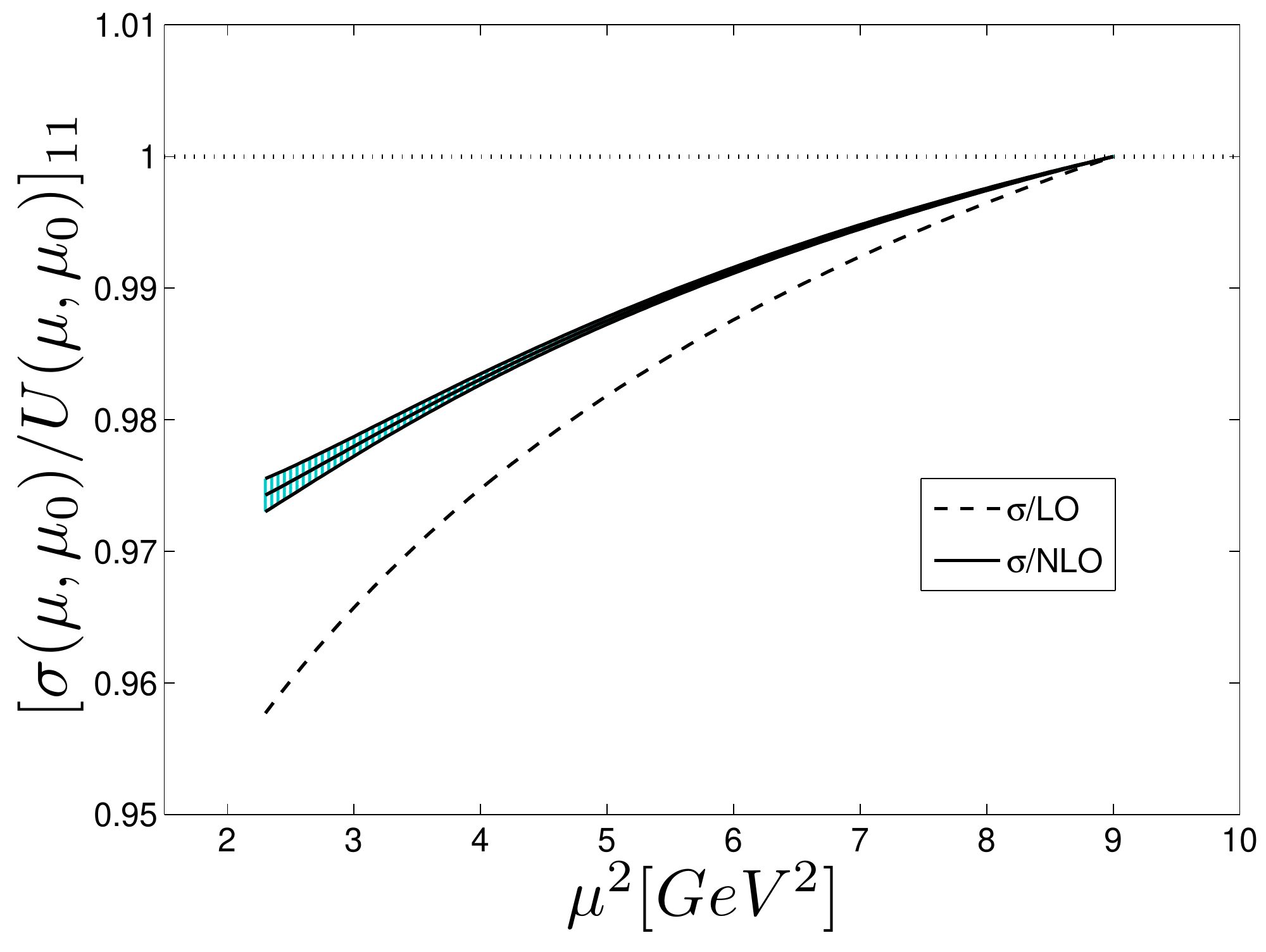}
\caption[]{
Step scaling function $\sigma_{11}(\mu,\mu_0)$ of the  $(27,1)$ operator 
divided by the corresponding perturbative running $U_{11}(\mu,\mu_0)$ 
(dashed line: one loop, solid coloured curve: two loops). For ease of comparison 
with perturbation theory and in contradiction to the previous plots,
$\mu_0$ is fixed at the conventional scale of $3\,\rm GeV$ and $\mu$
varies in the range $\left[1.5 \,\GeV,\, \mu_0 \right]$.
}
\label{fig:sigma11_over_pt}
\end{figure}
\begin{figure}[!h]
\begin{tabular}{cc}
\includegraphics[width=8.5cm]{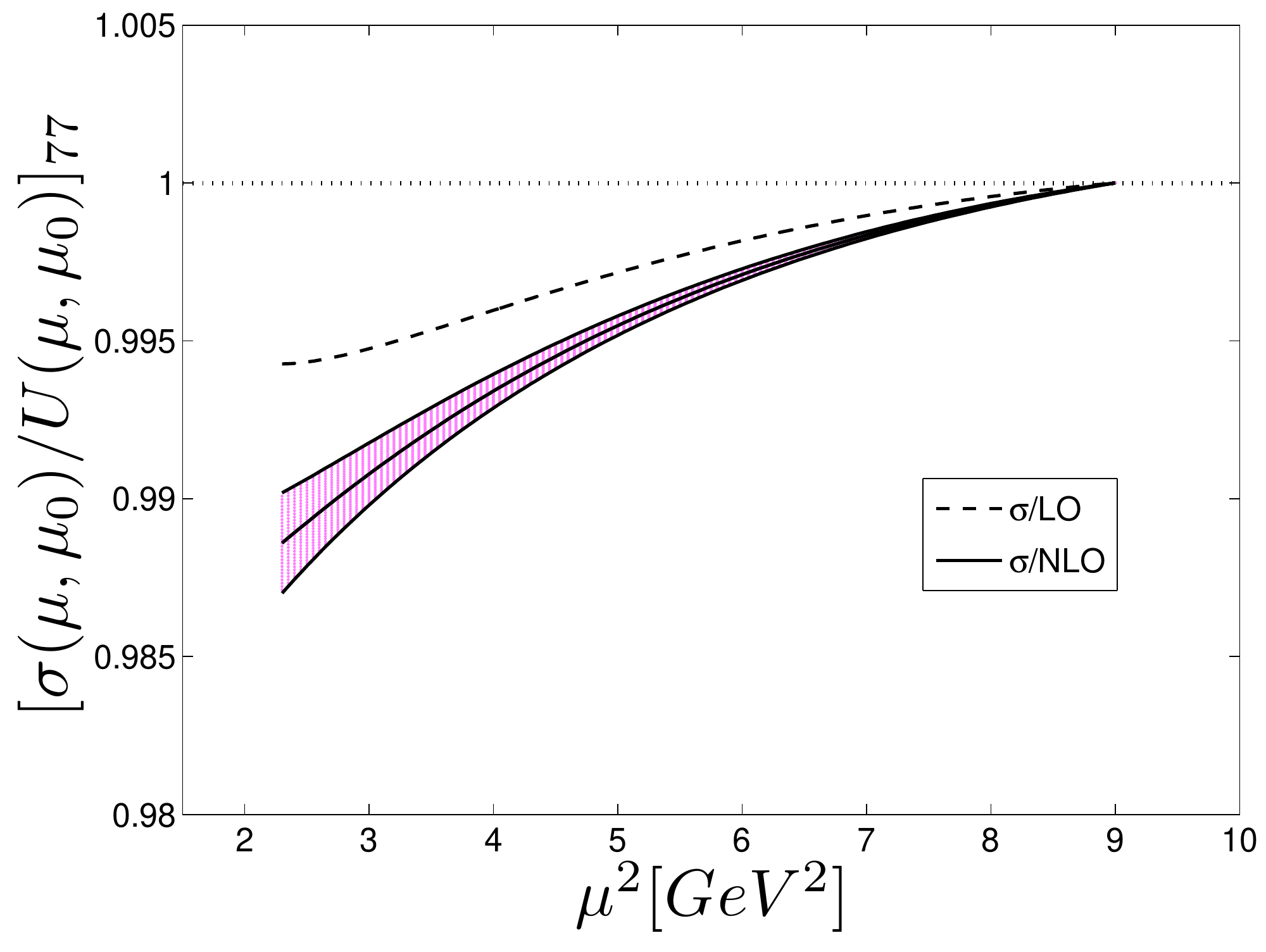}&
\includegraphics[width=8.5cm]{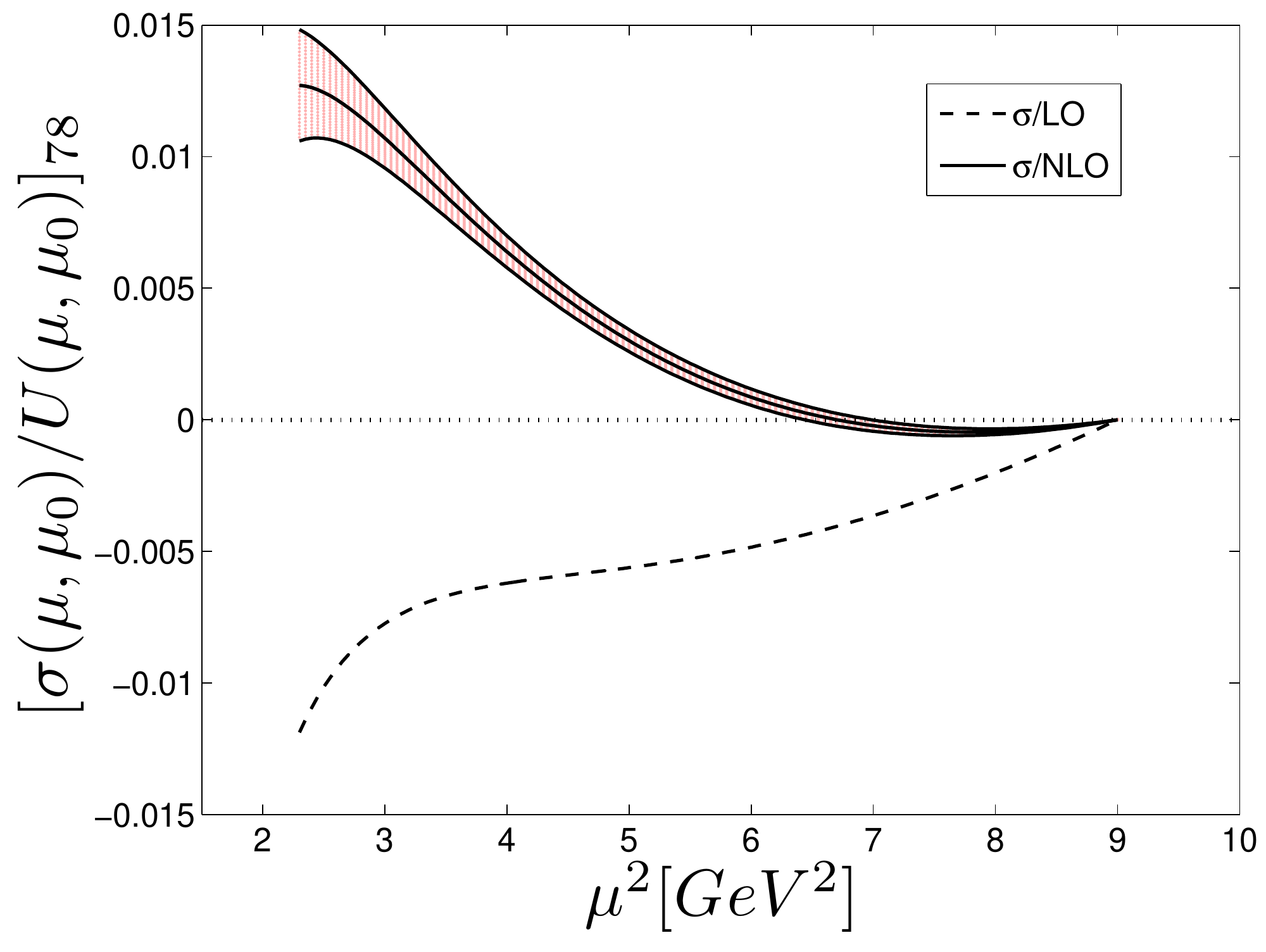}\\
\includegraphics[width=8.5cm]{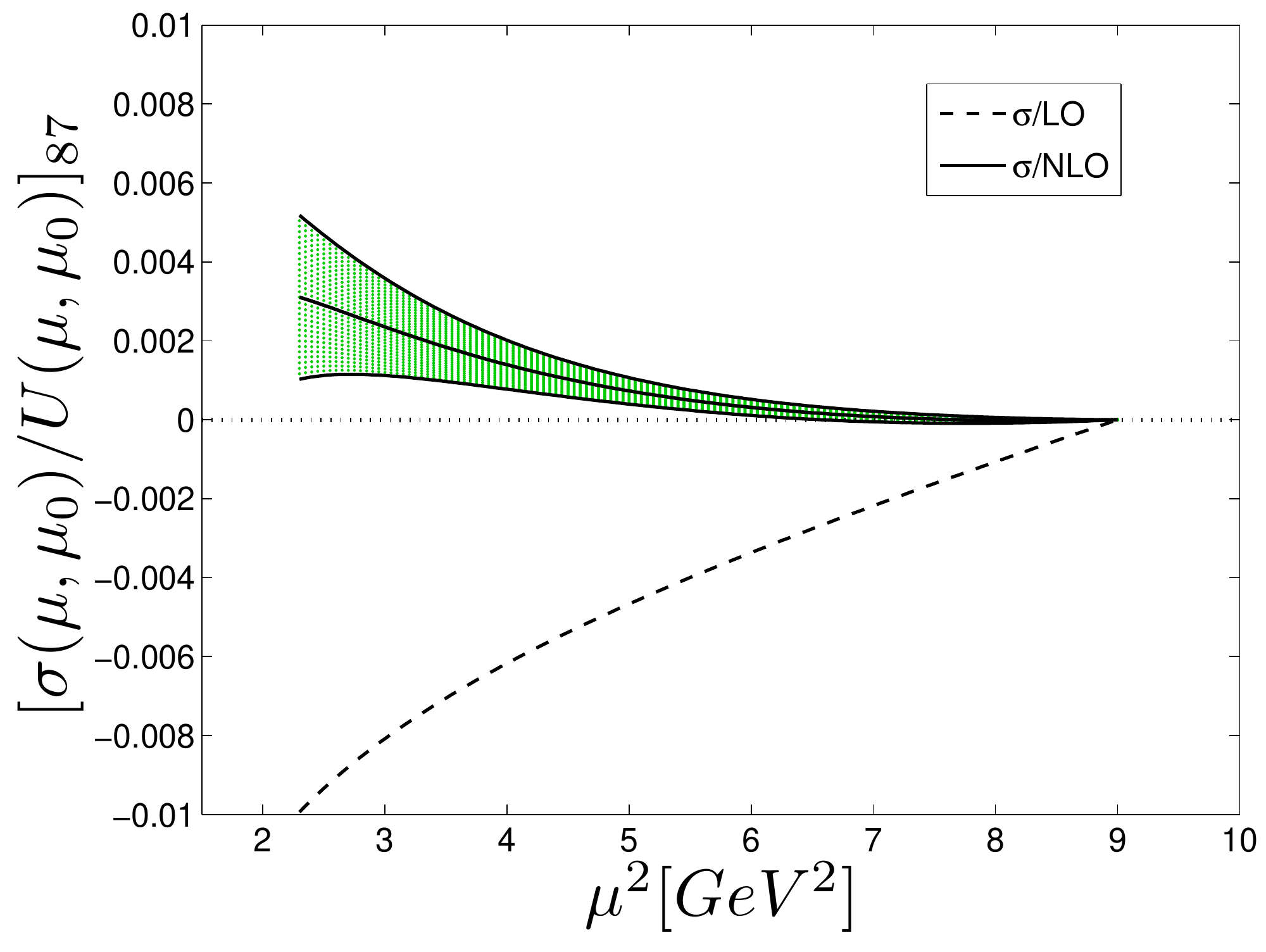}&
\includegraphics[width=8.5cm]{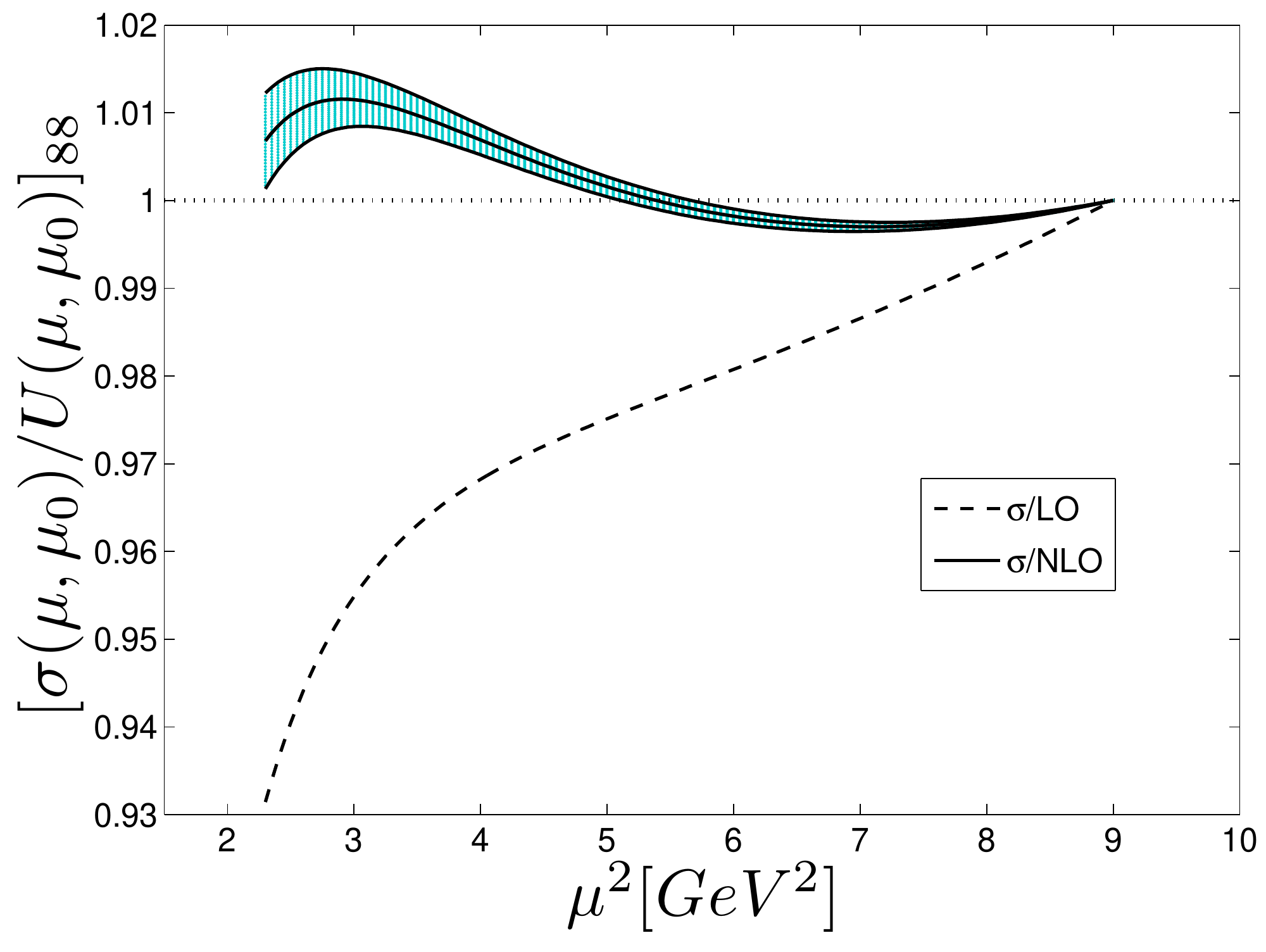}
\end{tabular}
\caption[]{
Same as figure \ref{fig:sigmaDs1} for the 
Step scaling matrix $\sigma(\mu,\mu_0)$ of the  $(8,8)$ electroweak penguins.
}
\label{fig:sigmaDs1_over_pt}
\end{figure}
\newpage
\end{widetext}

\bibliography{paper}{}
\bibliographystyle{h-elsevier}

\end{document}